\documentclass{aa}
\usepackage{graphicx}
\usepackage{lscape}
\usepackage{longtable}
\usepackage{txfonts}
\usepackage{natbib}

\begin{document}

   \title{Detection and high-resolution spectroscopy of a huge flare on 
the old M9 dwarf DENIS 104814.7-395606.1 \thanks{Based on observations
collected at the European Southern Observatory, Paranal, Chile, 68.D-0166A.}}

   \titlerunning{Flare on DENIS 104814.7-395606.1}

   \author{B. Fuhrmeister \and
          J. H. M. M. Schmitt 
          }

   \offprints{B. Fuhrmeister}

   \institute{Hamburger Sternwarte, University of Hamburg,
              Gojenbergsweg 112, D-21029 Hamburg\\
              \email{bfuhrmeister@hs.uni-hamburg.de}
             }


   \abstract{We report a flare on the M9 dwarf DENIS 104814.7-395606.1, whose mass places
             it directly at the hydrogen burning limit. The event was
             observed in a spectral sequence during 1.3 hours. Line shifts to
             bluer wavelengths were detected
             in $\mathrm{H_{\alpha}}$, $\mathrm{H_{\beta}}$, and in the Na\,{\sc i} D
             lines, indicating mass motions. In addition we detect  
             a flux enhancement on the blue side of the
             two Balmer lines in the last spectrum of our series. We interpret this as rising gas cloud
             with a projected velocity of about 100 $\mathrm{km s^{-1}}$ which may
             lead to mass ejection. The higher Balmer lines $\mathrm{H_{\gamma}}$ to
             $\mathrm{H_{8}}$ are not seen due to our instrumental setup, but in the last spectrum
             there is strong evidence for $\mathrm{H_{9}}$ being in emission.

   }

   \maketitle
%

\section{Introduction}
DENIS 104814.7-395606.1 (hereafter DENIS 1048-39) was discovered in the 
DEep Near Infrared Survey (DENIS), which covered the southern sky \citep{Epchtein} in two 
near infrared bands (J and $\mathrm{K_{s}}$) and one optical band (I).  With this choice of 
bands DENIS is very sensitive to very low mass stars and brown dwarfs and thus excellently suited
for searches for hitherto unknown low mass stellar or substellar objects in the solar vicinity. 
With a distance of only $4.6\pm0.3$\,pc \citep{DENISinfr} DENIS 1048-39 is extremely 
close to the Sun.
Classified as a M9 star, the lithium resonance line at \mbox{6708 \AA} could not be detected,
therefore  substantial lithium depletion must have taken place \citep{DENISdelfosse}. 
Compared to LP~944-20, which is also classified as M9, DENIS~1048-39 should be older and 
more massive. Theoretical
models place DENIS 1048-39 directly at the hydrogen burning boundary with an
estimated mass of 0.075 up to 0.09 $\mathrm{M_{Sun}}$ and an age of 1 - 2 Gyrs \citep{DENISinfr} 
assuming a solar chemical composition. Thus DENIS 1048-39 may either be among the most 
massive brown dwarfs or among the least massive stars. 

Interestingly, the $\mathrm{H_{\alpha}}$-line was found to be variable \citep{DENISdelfosse}, 
implying that DENIS 1048-39 exhibits activity despite
its old age and low mass. There are only a few very late type objects showing strong and persistent 
H$_{\alpha}$ emission.  H$_{\alpha}$ emission, a well established activity indicator, 
is not detected in the majority of the stars.
Instead one observes a steep decline of the strength of 
H$_{\alpha}$ emission for stars 
later than spectral type M7. This effect shows up in  smaller equivalent widths (EW) of the 
H$_{\alpha}$ line for these objects. However, since the
H$_{\alpha}$ line is seen against an increasingly faint photosphere for later objects, a better
activity indicator is the ratio of the H$_{\alpha}$ luminosity to the bolometric luminosity.
This ratio was found to drop in only three subclasses (M8-L1) by one order of magnitude 
\citep{Gizis}. The same authors also
found that the activity of these late type objects is primarily related to temperature and 
shows only a secondary dependence on mass and age. Thus strong H$_{\alpha}$ emission does not necessarily 
imply youth; on the contrary, strong H$_{\alpha}$ emitters in the field are more likely to 
be old. 

Despite this decline in activity there are many reports of flare activity even among L dwarfs.
For very late-type M dwarfs \citet{Gizis} estimated a flaring time fraction of about 7 \%, while
\citet{Liebert1} found a flare duty cycle of about 1 \% for L dwarfs, suggesting
that there must be some ongoing magnetic activity in these dwarfs despite the absence of a 
persistent chromosphere or corona.

In this paper we report on a huge flare on DENIS 1048-39 detected in the H$_{\alpha}$,  
H$_{\beta}$, and sodium emission lines.  
In section 2 we describe the VLT data and their analysis, in section 3 we discuss the 
timing behaviour of the flare.


\section{Observations and data analysis}

The observations reported in this paper were carried out 
with ESO's Kueyen telescope 
at Paranal equipped with the Ultraviolet-Visual Echelle Spectrograph (UVES) on
March, 14th, 2002. Four spectra of DENIS 1048-39 were taken in sequence, each with an 
exposure time of 1200 seconds. The individual observations started at 04:12:46, 04:33:43,
04:54:40 and 05:22:40 UT.
  The instrument was operated in dichroic mode, yielding
33 echelle orders in the blue arm (spectral coverage from 3030 to 3880~\AA) and 39 orders
in the red arm (spectral coverage from 4580 to 6680 \AA) with a typical resolution of $\sim45000$.
 The red part of the
spectrum was recorded on two separate CCDs; therefore there is a spectral gap from 
$\sim$ 5640 to 5740 \AA\, resulting from the spatial separation of the CCDs. As a consequence of the dichroic mode used
we did not cover the lines from H$_{3}$ up to H$_{8}$ of the Balmer series,
nor the Ca\,{\sc ii} H and K lines.  




The data were reduced using IRAF. 
The wavelength calibration was carried out with Thorium-Argon spectra with
an accuracy of $\sim 0.03$\AA\, in the blue arm and $\sim 0.05$\AA\, in the red arm.

The fitting of the spectral lines was carried out with the CORA fit program \citep{cora}, kindly provided
to us by Dr. Jan-Uwe Ness and
originally developed for analyzing high resolution X-ray spectra. The fit algorithms employed
by CORA are also well suited for a modeling of all types of emission lines with  count 
statistics. CORA
uses a maximum likelihood method for the fitting of line profiles.
It contains a number of tools for modeling blended lines, allowing the
user to specify the number of lines to be modeled, whether the line
shapes should be individually fitted or fitted with a fixed line profile,
and whether the line centroids are allowed to vary freely or are fixed
with respect to each other. The program also provides an accurate error analysis. 


\section{The UVES spectrum of DENIS 1048-39}

\subsection{Emission line variability}

\begin{figure}
\begin{center}
\includegraphics[width=9cm,height=5cm,clip=0,bbllx=100,bblly=100,bburx=560,bbury=380]{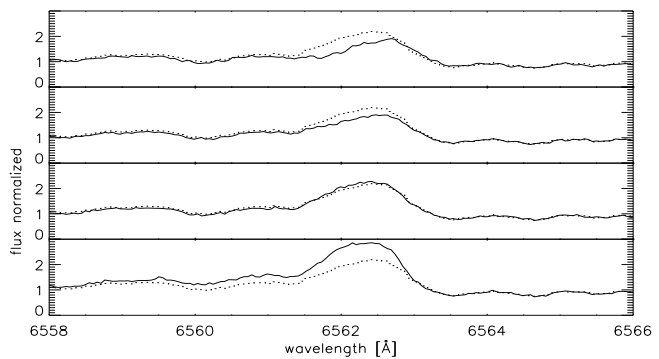}
\caption{\label{DENIS-Halpha}Time series of the $\mathrm{H_{\alpha}}$-line first spectrum
at the top, last spectrum at the bottom. The dotted spectrum is the time averaged spectrum
of DENIS 1048-39. Notice the gradual increase and blue shift of the line; in the last spectrum
a broad feature is seen to emerge on the blue side of the line centered on about 6560.5 \AA.}
\end{center}
\end{figure}

\begin{figure}
\begin{center}
\includegraphics[width=9cm,height=5cm,clip=0,bbllx=100,bblly=100,bburx=560,bbury=380]{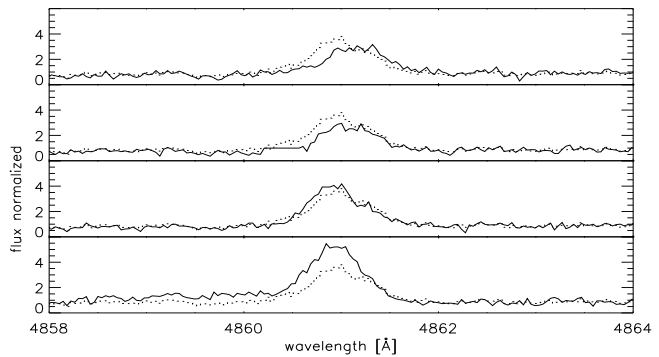}
\caption{\label{DENIS-Hbeta}Time series of the $\mathrm{H_{\beta}}$-line as in Fig. \ref{DENIS-Halpha}. 
Again a blue shift of the line can be seen as well as a broad emission feature
at the blue side of the line in the last spectrum. In the spectra some cosmics were replaced
manually by a horizontal line. The dotted line is again the time averaged spectrum.}
\end{center}
\end{figure}

\begin{figure}
\begin{center}
\includegraphics[width=9cm,height=5cm,clip=0,bbllx=100,bblly=100,bburx=560,bbury=380]{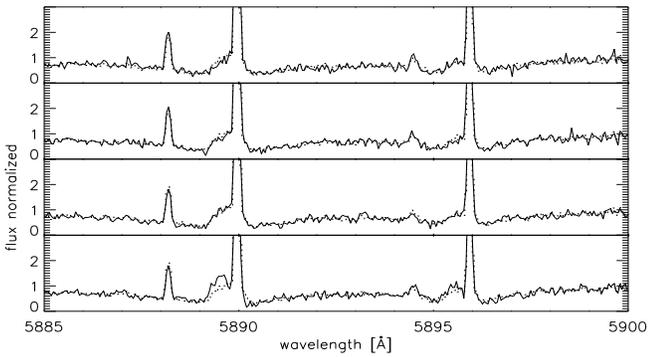}
\caption{\label{DENIS-NaD} Time series of the Na\,{\sc i} D lines as in Fig. \ref{DENIS-Halpha}. 
The dotted line denotes again the
time averaged spectrum. Note the blue shift that can be seen in comparison
to the airglow Na\,{\sc i} D lines especially in the last spectrum, where the peak of the stars Na\,{\sc i} D lines
moves out of the airglow lines.}
\end{center}
\end{figure}

In our UVES spectra of DENIS 1048-39 we find three prominent emission lines in
the red part of the spectra, which can be clearly identified as $\mathrm{H_{\alpha}}$,
$\mathrm{H_{\beta}}$, and the Na\,{\sc i} D doublet at 5889 and 5895 \AA. 
In Fig. \ref{DENIS-Halpha} - \ref{DENIS-NaD} we plot the spectral region around 
$\mathrm{H_{\alpha}}$, $\mathrm{H_{\beta}}$, and the Na\,{\sc i} D for the four individual spectra.
In order to facilitate the comparison between the spectra we also plot the 
averaged spectrum as a dotted line. Note that the spectra are not (yet) 
shifted to the star's rest frame.

In the spectral sequence of our VLT observations a clear intensity increase 
as well as a blue shift of the $\mathrm{H_{\alpha}}$ and $\mathrm{H_{\beta}}$ lines is seen.  In the last spectrum of our series an extremely broad emission feature appears, centered at 
about 6560.5 \AA\ on the blue side of the $\mathrm{H_{\alpha}}$ line.
A similar emission feature appears for the $\mathrm{H_{\beta}}$ line centered at about 4860 \AA. 
If interpreted as Doppler shift, this would 
imply emission from material moving with a velocity of 100 $\mathrm{km\,s^{-1}}$ 
towards the observer.

For the Na\,{\sc i} D lines the flux is not constant, but no clear trend can be seen. The  stellar Na\,{\sc i} D lines are heavily
blended with terrestrial airglow lines of Na\,{\sc i} D; in addition, there are further airglow lines in this spectral region at 5888.19 \AA\, 
and 5894.47 \AA, well known to occur in the UVES instrument \citep{Hanuschi}. The blue shift of the stellar Na\,{\sc i} D lines can be
unambiguously seen, when the stellar Na\,{\sc i} D line peak moves out of 
the stationary airglow lines in the last spectrum. We therefore conclude that the recorded line shifts cannot be due to incorrect wavelength calibrations.   

\begin{figure}
\begin{center}
\includegraphics[width=9cm,height=5cm,clip=0,bbllx=100,bblly=100,bburx=560,bbury=380]{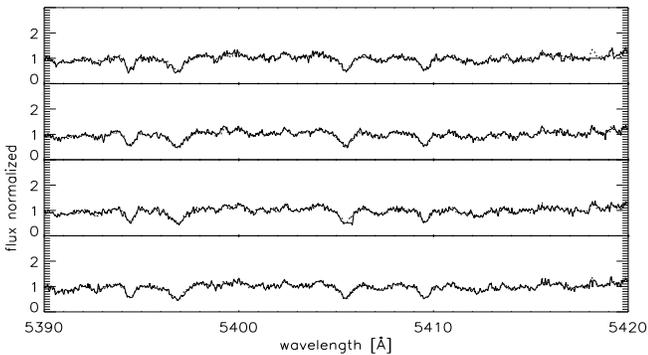}
\caption{\label{absorptionlines} Timeseries of a spectral range with several metal absorption lines.
Clearly these photospheric lines are not affected by the blue shift seen in the chromospheric
emission lines. Two cosmics were removed manually by a horizontal line. The four pronounced absorption
lines are due to Mn\,{\sc i} (5394 \AA), Fe\,{\sc i} (5397 and 5405 \AA) and Cr\,{\sc i}  
(5409 \AA). Again the dotted line denotes the time averaged spectrum.}
\end{center}
\end{figure}

In Fig. \ref{absorptionlines}
we plot the time series of the spectrum between 5390 \AA - 5420 \AA,
which is purely of photospheric origin.  This spectral region obviously remains constant and thus the photospheric continuum appears to 
be unaffected by the flaring emission lines.

\subsection{Radial and rotational velocity}

Having confirmed that the absorption lines are not affected by any blue shifts, we measured the
radial velocity of the star to be $-9.6 \pm 2.0\,\mathrm{km\,s^{-1}}$ in agreement with the
value of $-10.1 \pm 0.5 \mathrm{km\,s^{-1}}$ found by \citet{DENISdelfosse}. All line fitting 
with CORA was done after shifting the spectra to the rest frame.
In addition to the radial velocity we also measured the rotational velocity of 
DENIS~1048-39 using 7 apertures in the 
red arm of the spectrum without any emission and airglow lines. We used the spectrum
of the M9 star LHS~2065 (taken during the same observation run) as a template and convolved it with rotational velocities from 3 up to 40 
$ \mathrm{km\,s^{-1}}$. The best rotational fit of the spun-up template and DENIS~1048-39
was determined with a $\chi^{2}$ test for every aperture used.  The values determined for the 
different apertures were then averaged to yield a final
(averaged) rotational velocity of 25.0 $\pm$ 2 $ \mathrm{km\,s^{-1}}$ in good
agreement with the $\mathrm{v}\sin(i)$-value of 27 $\pm$ 5 $ \mathrm{km\,s^{-1}}$
found by \citet{DENISdelfosse}.


\subsection{The Na\,{\sc i} D lines}

We fitted the stellar and airglow lines of Na\,{\sc i} D simultaneously
treating the central wavelength $\lambda_{cen}$, Gaussian $\sigma$ and amplitude for every line 
as free parameters.  The fit results are listed in Table \ref{NaD}. While the central wavelength 
of the airglow lines is stable as it should, our fits indicate a steady bluewards drift in the 
stellar Na\,{\sc i} D lines. The line shifts were computed using 5889.950
and 5895.924 \AA\, as reference central wavelengths; although the errors are large, the trend to 
bluer line centers is clearly visible. 

The large amplitude for the line at 5895.924 \AA\, in the first spectrum does not agree with the line amplitudes of the other three spectra that vary only slightly. We ascribe this to
the heavy blending of the line and the faintness of the feature which makes its amplitude
sensitive to noise and to the background level chosen for the fit.

\begin{table*}
\caption{\label{NaD}Properties of the Na\,{\sc i} D lines determined from
 fitting the Na\,{\sc i} D airglow line and the
star's Na\,{\sc i} D line simultaneously. Quoted are all free fit parameters and the line shift of the star's
Na\,{\sc i} D line. Due to the heavy blending the error estimation of the
half width could not be done for each spectrum.}
\begin{center}
\begin{tabular}[htbp]{cccccc}
\hline
             &                & star & airglow & star & airglow\\
\hline
1st spectrum & $\lambda_{cen}$ [\AA] & 5890.00 $\pm$ 0.04& 5890.19$\pm$ 0.02& 5895.96 $\pm$ 0.03 & 5896.17 $\pm$ 0.02 \\
             & $\sigma$ [\AA] &0.30 $\pm$ 0.03& 0.05 $\pm$ 0.01& 0.25  & 0.05 $\pm$ 0.01\\
             & amplitude [electrons] & 1920 $\pm$ 73& 5890 $\pm$ 85 & 1225 $\pm$ 65 & 3665 $\pm$ 69 \\
             & line shift [$\mathrm{km\ s^{-1}}$]& 2.5 $\pm$ 2.0 &  & 1.8 $\pm$ 1.1&  \\
\hline
2nd spectrum & $\lambda_{cen}$ [\AA] & 5889.97 $\pm$ 0.03& 5890.20 $\pm$ 0.03& 5895.91 $\pm$ 0.04& 5896.16 $\pm$ 0.02\\
             & $\sigma$ [\AA]& 0.28 $\pm$ 0.01 & 0.06 $\pm$ 0.01 & 0.24 $\pm$ 0.02 & 0.06 $\pm$ 0.01 \\
             & amplitude [electrons] & 2110 $\pm$ 74 & 5258 $\pm$ 82& 942 $\pm$ 60 & 3359 $\pm$ 66  \\
             & line shift [$\mathrm{km\ s^{-1}}$]& 1.0 $\pm$ 1.5 &   & $-$0.71 $\pm$ 2.0 &     \\
\hline
3rd spectrum & $\lambda_{cen}$ [\AA]& 5889.92 $\pm$ 0.02& 5890.20 $\pm$ 0.01& 5895.86 $\pm$ 0.05& 5896.17 $\pm$ 0.01\\
             & $\sigma$ [\AA]& 0.24 $\pm$ 0.01 & 0.06 $\pm$ 0.01& 0.19 & 0.06 $\pm$ 0.01 \\
             & amplitude [electrons] & 2316 $\pm$ 73 & 5208 $\pm$ 81 & 811 $\pm$ 54  & 3475 $\pm$ 66 \\
             & line shift [$\mathrm{km\ s^{-1}}$]& $-$1.5 $\pm$ 1.0 &  & $-$3.3 $\pm$ 2.5 &    \\
\hline
4th spectrum & $\lambda_{cen}$ [\AA]& 5889.76 $\pm$ 0.04& 5890.19 $\pm$ 0.02& 5895.78 $\pm$ 0.07  & 5896.17 $\pm$ 0.02  \\
             & $\sigma$ [\AA] & 0.19 $\pm$ 0.01 & 0.06 $\pm$ 0.01& 0.14 $\pm$ 0.01 & 0.05 $\pm$ 0.01 \\
             & amplitude [electrons]& 2273 $\pm$ 62 & 5069 $\pm$ 76 & 832 $\pm$ 47 & 2977 $\pm$ 60\\
             & line shift [$\mathrm{km\ s^{-1}}$]& $-$9.7 $\pm$ 2.0 &  & $-$7.3 $\pm$ 3.6&  \\
\hline
\end{tabular}
\end{center}
\end{table*}


\subsection{The $\mathrm{H_{\alpha}}$  and the $\mathrm{H_{\beta}}$ line}

\begin{figure}
\begin{center}
\includegraphics[width=8cm,height=5cm,clip=0]{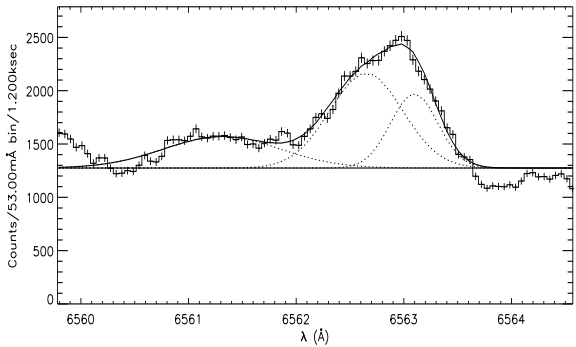}
\caption{\label{Halpha1}Fit of the $\mathrm{H_{\alpha}}$ line in the first spectrum of the
series. Two fit components are used for the line, while the third component accounts for the
bump in the continuum blue to the line. The horizontal line is the background used for the fit.
The dotted lines mark the single gaussian components used for the fit. Their free fit parameters
can be found in Table \ref{balmer}.}
\end{center}
\end{figure}

The amplitudes of the $\mathrm{H_{\alpha}}$ and $\mathrm{H_{\beta}}$ lines were fitted with a single Gaussian line profile as for the Na\,{\sc i} D lines. 
For the $\mathrm{H_{\beta}}$ line this seems to be a reasonably good 
description although the line center shows some fine structure. However, for the
$\mathrm{H_{\alpha}}$ line these fits yielded only a very poor description of line profiles in UVES data. 
A major problem for our fits of the $\mathrm{H_{\alpha}}$ line is 
the background treatment, which is somewhat complicated with several unidentified
broad emission and absorption features on the blue side
of the line and a slightly higher background level on the blue compared to the red side.
Although there appears to be some variation in the background level 
we decided to use a constant background since there is no other reasonable method to treat the
background problem i.\,e., there is no obvious  gradient. We used the 
background value redwards of the
line found in the second spectrum for our fits. 
Therefore the first broad emission feature at 6561.29 \AA\,
(see Fig. \ref{Halpha1}) must be treated as an additional line feature
in the fit of the $\mathrm{H_{\alpha}}$ line.  
The $\mathrm{H_{\alpha}}$ line also shows a strong asymmetry in 
the blue wing, where more and more flux appears during our spectral sequence, while the red 
wing of the line stays almost constant. We therefore decided to use two spectral components 
to describe the shape of the line core and a
third component to account for the closest blue emission feature in the background. 
All fit components were assumed to be Gaussians and we use again central wavelength 
$\lambda_{cen}$, Gaussian $\sigma$ and the line amplitude
as free fit parameters. 
The results of our fits of the $\mathrm{H_{\alpha}}$  and the $\mathrm{H_{\beta}}$ line are listed in Table \ref{balmer}.  
In order to provide an example of the quality of our fits we show 
the fitted $\mathrm{H_{\alpha}}$ line for the first spectrum in Fig. \ref{Halpha1}. While
not necessarily physical, the three components do obviously provide a good
analytical description of the recorded UVES spectrum.

\begin{figure}
\begin{center}
\includegraphics[width=8cm,height=5cm,clip=0]{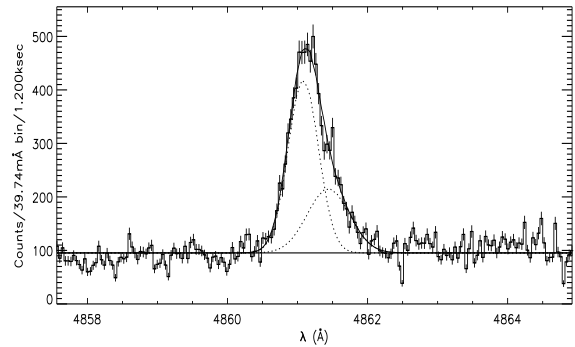}
\caption{\label{Hbeta}Fit of the $\mathrm{H_{\beta}}$ line in the third spectrum of the
series. Two fit components are used for the line resulting in a good fit of the line especially
in the red wing. The horizontal line is the background used for the fit.
The dotted lines mark the single Gaussian components. }
\end{center}
\end{figure}

Since two Gaussian components are required for the fit of the core of the $\mathrm{H_{\alpha}}$ line, the question arises, why the $\mathrm{H_{\beta}}$ line behaves differently and can be fitted with a single component. Indeed, a close inspection of the line shape of the $\mathrm{H_{\beta}}$ line 
in the third spectrum also shows an additional red component. Therefore we tried to fit the $\mathrm{H_{\beta}}$ line with two components as well. For the third spectrum this leads to a substantial improvement of the fit resulting in similar line shifts and a similar ratio of the two
components as for $\mathrm{H_{\alpha}}$ (see Fig. \ref{Hbeta}); unfortunately for the other spectra a fit with two components describing the line core does not lead to unique solutions and does not provide significantly improved fit results. Therefore we decided to use the fits
with one component for $\mathrm{H_{\beta}}$ bearing in mind that there is evidence for a second
red component of the line that behaves similarly to the red component of the $\mathrm{H_{\alpha}}$ line
at least for the third spectrum.   

Our fit results for the $\mathrm{H_{\alpha}}$ line clearly demonstrate 
that the reddest line component is only weakly blue shifted during our observations, while the blue
component is shifted by 12.3 $\mathrm{km\,s^{-1}}$ bluewards from its initial position during 
the first three spectra. In the last spectrum there are no additional line shifts observed 
neither for the $\mathrm{H_{\alpha}}$ line components nor for the $\mathrm{H_{\beta}}$ line, rather a substantial
flux increase occurs in the blue component of the $\mathrm{H_{\alpha}}$ line and in the $\mathrm{H_{\beta}}$ line. For the $\mathrm{H_{\alpha}}$ line the ratio between the amplitude of the two components
is approximately constant for the first three spectra, but in the last spectrum the blue 
component is observed with some excess flux. We also note that the line shifts measured for 
the $\mathrm{H_{\beta}}$ line are consistent with the
line shifts found in the Na\,{\sc i} D lines.  

\begin{table*}
\caption{\label{balmer}Properties of the $\mathrm{H_{\alpha}}$  and the $\mathrm{H_{\beta}}$ line.
The line shift is given in $\mathrm{km\ s^{-1}}$ using 4861.332 and 6562.817 \AA\, as
reference central wavelength.}
\begin{center}
\begin{tabular}[htbp]{ccccccc}
\hline
             &                & \multicolumn{4}{c}{$\mathrm{H_{\alpha}}$} & $\mathrm{H_{\beta}}$ \\
             &                & bump & 1st comp& 2nd comp& total &       \\
\hline
1st spectrum & $\lambda_{cen}$ [\AA] & 6561.29 $\pm$ 0.04& 6562.62 $\pm$ 0.01 & 6563.07 $\pm$ 0.01 && 4861.36 $\pm$ 0.12\\  
             & $\sigma$ [\AA] & 0.54 $\pm$ 0.04 & 0.34$\pm$ 0.01 & 0.22$\pm$0.01&& 0.31 $\pm$ 0.02\\
             & amplitude [electrons] & 7477.3$\pm$237.2 & 14094.2$\pm$259.0& 7357.9 $\pm$ 211.0&21452 $\pm$ 334& 4032 $\pm$ 81\\
             &line shift [$\mathrm{km\ s^{-1}}$] &                  &$-$9.0$\pm$0.5&11.6$\pm$0.5&& 1.7$\pm$7.4\\
\hline
2nd spectrum & $\lambda_{cen}$ [\AA]& 6561.19$\pm$ 0.05& 6562.40$\pm$0.01& 6562.96$\pm$0.01&& 4861.29 $\pm$ 0.10\\
             & $\sigma$ [\AA] & 0.40$\pm$0.01& 0.42$\pm$0.01&0.25$\pm$0.01&& 0.26 $\pm$ 0.02\\
             & amplitude [electrons] & 8003.8$\pm$219.3& 22086.6$\pm$301.1& 10642.6$\pm$232.5&32728$\pm$380& 3730 $\pm$ 78\\
             &line shift [$\mathrm{km\ s^{-1}}$]  &                  & $-$19.0$\pm$0.5 & 6.5$\pm$0.5 &&$-$2.6$\pm$6.1\\
\hline
3rd spectrum & $\lambda_{cen}$ [\AA] & 6561.13$\pm$0.05&6562.36$\pm$0.01&6562.92$\pm$0.01&& 4861.15$\pm$0.10\\
             & $\sigma$ [\AA] & 0.36$\pm$0.01& 0.40$\pm$0.01& 0.26$\pm$0.01&& 0.28$\pm$0.01\\
             & amplitude [electrons] & 7669.0$\pm$213.6& 33246.9$\pm$325.3& 15394.6$\pm$261.5&48642 $\pm$ 416& 6460.9$\pm$97.2\\
             &line shift [$\mathrm{km\ s^{-1}}$] &                  & $-$21.3$\pm$0.5 & 4.7$\pm$0.5 &&$-$11.2$\pm$6.1\\
\hline
4th spectrum & $\lambda_{cen}$ [\AA] & 6561.19$\pm$0.01&6562.39$\pm$0.01&6562.93$\pm$0.01&& 4861.14$\pm$0.02\\
             & $\sigma$ [\AA] & 0.54$\pm$0.01& 0.36$\pm$0.01&0.22$\pm$0.01&& 0.26$\pm$0.01\\
             & amplitude [electrons]& 22450$\pm$288.5& 43569.8$\pm$340.2&16920$\pm$253.5&60490 $\pm$ 423& 7841.0$\pm$107.1\\
             &line shift [$\mathrm{km\ s^{-1}}$] &                  & $-$19.4$\pm$0.5 & 4.7$\pm$0.5  &&$-$11.8$\pm$1.2\\
\hline
\end{tabular}
\end{center}
\end{table*}

\begin{figure}
\begin{center}
\includegraphics[width=8cm,height=5cm,clip=0,bbllx=25,bblly=20,bburx=610,bbury=300]{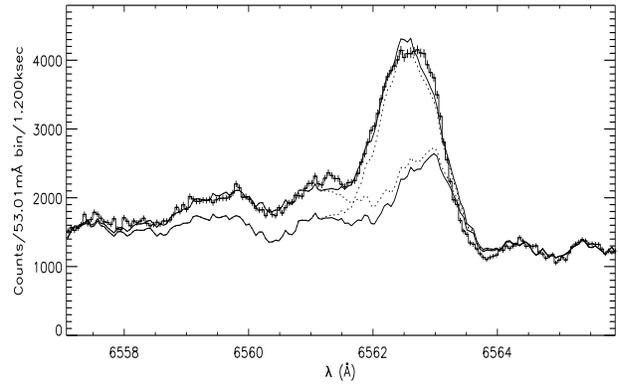}
\caption{\label{Halphabump} Fit of the broad emission feature blueward of the $\mathrm{H_{\alpha}}$ line
in the last spectrum of the series.
The lower black line is the first spectrum of the series that was used as background. Note the flux increase
between the two spectra not only in the line core but also in the broad emission feature centered at
6560.5 \AA\, blueward of the
line. The two dotted lines
are the single gaussian components used for the fit. Since the background is variable their shape does
not appear to be gaussian. Their
free fit parameters can be found in Table \ref{broademission}.
}
\end{center}
\end{figure}

In the last spectrum of our spectral series
a tremendous flux increase occurs bluewards of the $\mathrm{H_{\alpha}}$ line. This increase 
can be seen in Fig. \ref{DENIS-Halpha}, but most clearly in Fig. \ref{Halphabump}.   
The ``background" on the blue and red side of the  $\mathrm{H_{\alpha}}$ line agrees for all 
spectra except for the blue side of the last spectrum in the spectral series.  There a flux is 
detectable out to
5 \AA\, from line center, while blueward of 6558 \AA\, the two background
levels are identical again. 

In order to fit this additional flux in the last spectrum
we used the first spectrum of the series as a background ``template". Obviously with this strategy we will not obtain meaningful results for the fit of the $\mathrm{H_{\alpha}}$ line itself, however, we merely intend to characterize
the flux increase on the blue side of the $\mathrm{H_{\alpha}}$ line.
For the fit we used two Gaussian components with the same free fit parameters
as before. Since the fit of the $\mathrm{H_{\alpha}}$ line core is ill defined due to the background
anyway we only used one Gaussian component for this core, while the other Gaussian is used to fit
the broad emission component (see Fig. \ref{Halphabump}). 

For the
$\mathrm{H_{\beta}}$ in the last spectrum there is a similar broad emission feature of the blue side
of the line as can be seen from Fig. \ref{DENIS-Hbeta}. 
In this case the surrounding background is quite flat, therefore a fit of this broad emission
line with a constant background would have been possible but for consistency
reasons we fitted the broad emission
feature with the first spectrum as background, too. 
The best fit parameters can be found in
Table \ref{broademission}.

\begin{table}
\caption{\label{broademission}Properties of the broad emission feature bluewards of the
$\mathrm{H_{\alpha}}$  and the $\mathrm{H_{\beta}}$ line in the last spectrum.
The line shift is given in $\mathrm{km\ s^{-1}}$ using 4861.332 and 6562.817 \AA\, as
reference central wavelength.}
\begin{center}
\begin{tabular}[htbp]{ccc}
\hline
                & $\mathrm{H_{\alpha}}$ & $\mathrm{H_{\beta}}$\\ 
\hline
$\lambda_{cen} [\AA]$ & 6560.51 $\pm$ 0.01&4859.76 $\pm$ 0.11\\
$\sigma$ [\AA] & 1.30 $\pm$ 0.01& 0.80 $\pm$ 0.03\\
amplitude [electrons] &27347.1 $\pm$ 437.1&3645.3 $\pm$ 104.7\\
line shift [$\mathrm{km\ s^{-1}}$] & $-$109.9 $\pm$ 0.5 & $-$96.9 $\pm$ 6.8\\

\hline
\end{tabular}
\end{center}
\end{table}

The line shifts of the broad emission features bluewards 
of the $\mathrm{H_{\alpha}}$ and the $\mathrm{H_{\beta}}$ line 
are not in agreement with each other due to the small error for the $\mathrm{H_{\alpha}}$ 
line (0.5 $\mathrm{km\ s^{-1}}$). In contrast
to this the error for the line shift of the broad component corresponding to the $\mathrm{H_{\beta}}$ line
is quite big (6.8 $\mathrm{km\ s^{-1}}$). Both errors should be considered only as formal ones since they
are only statistical errors. The unusual
background treatment can lead to additional systematic errors.   
If the errors for the $\mathrm{H_{\alpha}}$ line are as large as that of the $\mathrm{H_{\beta}}$ line, 
the line shifts actually agree very well.

Since absolute flux measurements were not the primary goal of this
observing run, we obtained only a few measurements of standard stars.
We calibrated our spectra with the standard star HD 49798, estimating the errors in the resulting 
fluxes to be as large as a factor of two. The measured line counts of the two main components of 
the $\mathrm{H_{\alpha}}$ line then translate into a flux of $6.3 \cdot
10^{-14}\, \mathrm{erg\ cm^{-2} s^{-1}}$, resulting in a line luminosity
of $1.6 \cdot 10^{26}\, \mathrm{erg\ s^{-1}}$ using a distance of 4.6 pc. This is not particularly 
large compared to other $\mathrm{H_{\alpha}}$ line luminosities
found for example for 2MASSW J0149090+295613 to be $15.3 \cdot 10^{26}\, \mathrm{erg\ s^{-1}}$ during a flare 
\citep{Liebert}. Estimating the bolometric luminosity with $M_{J}=11.28$ mag
\citep{DENISinfr} and BC = 1.95 mag leads to $L_{\mathrm{bol}}=1.5 \cdot 10^{30}\, \mathrm{erg\ s^{-1}}$ which yields  
a ratio of log($L_{\mathrm{H}\alpha}$ / $L_{\mathrm{bol}}$)=$-$4.0. According
to \citet{Liebert1} this would be quite high for an M9.5 dwarf in quiescent emission, but it
is well in the range of observed $L_{\mathrm{H}\alpha}$ to $L_{\mathrm{bol}}$ ratios for the
few flaring events known for objects at the M to L spectral class transition.


\subsection{Other Balmer lines}
\begin{figure}
\begin{center}
\includegraphics[width=8cm,height=5cm,clip=0,bbllx=25,bblly=20,bburx=610,bbury=300]{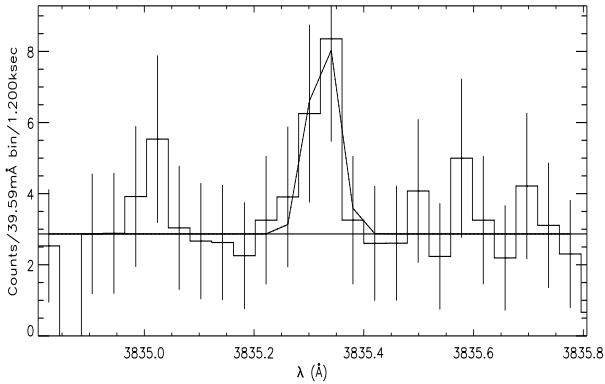}
\caption{\label{H9} The last spectrum in the series around 3835.386 \AA. The fit shows the
faint emission line of H$_{9}$. The horizontal line is the background used for the fit.}
\end{center}
\end{figure}

Due to the chosen instrumental setup we do not cover the H$_{3}$ up to H$_{8}$ of the Balmer series. 
Therefore we searched for emission lines at the position of H$_{9}$ and higher Balmer series members. 
Unfortunately the blue part of the spectra is very noisy since the count rate is very low. 
We do find a faint emission line at 3835.31 $\pm$ 0.04 \AA\, in our last
spectrum, corresponding to a blue shift of $-$5.9 $\pm$ 3.1 $\mathrm{km\ s^{-1}}$ which is less than that of the $\mathrm{H_{\beta}}$ line. Since $\mathrm{H_{\alpha}}$ and 
$\mathrm{H_{\beta}}$ differ significantly in appearance and blueshift, an interpretation of
this faint line as H$_{9}$ is very attractive.  We found the best
fit with an amplitude of 9.1 $\pm$ 4.0 electrons, $\sigma$ of 0.02 $\pm$ 0.01 \AA\, 
 and central wavelength of 3835.31 $\pm$ 0.04 \AA. The line fit can be seen in Fig. \ref{H9}. We also
searched for higher Balmer lines. In the last spectrum
an emission feature 
at 3797.76 $\pm$ 0.12 \AA\, corresponding to a blue shift of $-$11.0 $\pm$ 9.5 $\mathrm{km\ s^{-1}}$
of H$_{10}$ is found, but
it is so weak that it can be due to noise. On the wavelength
positions of H$_{11}$ no emission is found.

\section{Discussion and conclusions}

\subsection{Emission geometry}

The measured half widths of the Na\,{\sc i} D and the H$_{\beta}$ lines always 
stay below 20 $\mathrm{km\,s^{-1}}$, i.e., they are
smaller than the measured rotational velocity. The same applies to the
two main components of the H$_{\alpha}$ line.
This suggests that the emission for the Balmer lines
as well as for the Na\,{\sc i} D lines is confined to a restricted region on the star. 
Since the H$_{\alpha}$ line consists of two components and there is evidence that the
H$_{\beta}$ line consists of two components as well, the star could host even two
active regions producing the observed Balmer emission. Now the question arises what the nature of these
two active regions may be. There are two main possibilities: a dynamic scenario or a static scenario.

\subsection{Static or dynamic scenario ?}

The dynamic scenario invokes two active regions located on the surface of the star and mass motions within these active regions.
The Balmer emission comes from the chromosphere, nevertheless very close to the surface of the star.  At least in one of the 
two regions a brightening in the Balmer lines
takes place on a timescale of about 1.5 hours. H$_{\alpha}$ and H$_{\beta}$ emitting material moves
towards the observer and causes the blue shift.

The static scenario interprets the two regions
as emitting gas clouds corotating in some distance to the stars surface as was first
proposed for the K0 star AB~Dor by \citet{Cameron}. Since for one component the line shift  
becomes bluer this one must be rising while the other one must be about to set behind the star
since its line shift is moving towards the blue. The blue shifted component must then be captured during
its rise above the horizon of the star to account for the flux increase.
This latter static scenario can be excluded because of
the rapid rotation of the stars as follows: For
an estimated  radius of $\mathrm{R_{\star}} = 0.1 \mathrm{R_{\odot}}$ the maximal rotation 
period of DENIS 1048-39 is 4.9 hours. Therefore in the four consecutive spectra lasting
together 80 minutes the star completes about a fourth of its rotation. If the emitting gas
is confined in a corotating cloud, the measured velocity shift should exceed the star's
rotational velocity in at least one spectrum 
for orbits near the equatorial plane. Moreover the flux increase cannot be explained by a cloud
rotating into view because the star rotates too fast for this interpretation. 
A more natural explanation is therefore the scenario of
active regions with mass motions at the surface of the star that brighten during the spectral series.

\subsection{Blue-shifted Balmer emission}

Let us now consider the broad emission features bluewards of the two Balmer lines seen only in the 
fourth spectrum. Again a dynamic or static scenario may be considered. Let us first assume a 
static interpretation with a corotating cloud.  Since its radial velocity of about 100 
$\mathrm{km\,s^{-1}}$ is too high to be interpreted by a region on or near the surface of the 
star, the emission would have to come from a cloud at some distance to the star. Such a cloud must
be confined then by 
magnetic loops above the surface of the star and has rotated
just into the field of view. If one assumes a stellar inclination close to $90^{\circ}$ and a
cloud in the equatorial plane, one can compute its distance $R$ from the rotation axis with
$R=\frac{\mathrm{v} P}{2\pi}$ with $\mathrm{v}$ denoting the radial velocity and $P$ the 
rotation period of DENIS~1048-39. 
We find a distance of $R=4 \mathrm{R_{\star}}$ with the measured radial velocity of 100 
$\mathrm{km\,s^{-1}}$, which is below the Keplerian corotation radius of $6 \mathrm{R_{\star}}$
like the clouds found on AB~Dor in $\mathrm{H_{\alpha}}$ \citep{Donati} and other chromospheric 
lines \citep{Brandt}. But since the 
star rotates fast, the feature should
be visible in more than one spectrum since for its distance of four stellar
radii it can be shadowed by the star only 
about 10 \% of the time, i.\,e. about 30 minutes, thus again a dynamic
interpretation appears more likely.

In the dynamical interpretation the line shift
is explained in terms of material ejected by the star. 
Since the emission in the main body of the 
$\mathrm{H_{\alpha}}$ brightens substantially in the same spectrum it is suggestive that
the broad emission feature is connected to the active region producing the 
$\mathrm{H_{\alpha}}$ line arguing further in favor of a dynamic scenario. 
In the context of interpreting the broad emission feature
as the signature of a rising cloud the question remains unsettled whether this rise leads to
mass ejection since we do not know the longitude of the cloud. Assuming again an radius of
$0.1 \mathrm{R_{\odot}}$ we find an escape velocity of about 550 $\mathrm{km\,s^{-1}}$ for a
mass between 0.075 and 0.09 $\mathrm{M_{\odot}}$ while the projected velocity of the cloud is about 100 $\mathrm{km\,s^{-1}}$. 

Let us now consider the width of the emission feature. This width can be
due to temperature and turbulent broadening if one thinks of a confined gas cloud rising
towards the observer. A first estimate of the temperature of a $\mathrm{H_{\alpha}}$ emitting 
gas is about 10\,000 K leading to a thermal broadening of 
9 $\mathrm{km\ s^{-1}}$, far less than the observed total line broadening  of
50 $\mathrm{km\ s^{-1}}$. In this scenario with a rising cloud one can estimate
the height of the cloud after the twenty minutes exposure. Assuming
a constant cloud velocity of the cloud and an ejection start right
at the beginning of the exposure leads to
a cloud height of 1.7 $\mathrm{R_{\star}}$. It is probable that the cloud has been
decelerated during such a rise. Therefore a second interpretation of the line width 
is that different velocities of the cloud during deceleration are integrated over
the exposure time. Moreover, the cloud may consist of more than one component with different velocities. 
  
Besides these uncertainties in the event geometry there is no doubt that a dynamical
interpretation of the spectra is needed. The last spectrum is then quite suggestively interpreted
as  the onset of a flare on DENIS~1048-39 as
reported for AD~Leo by \citet{Houdebine} who found a similar flux enhancement in the far blue wing
of the $\mathrm{H_{\gamma}}$ line during a particular violent flare on AD~Leo. Since these authors
found projected velocities of up to 5800 $\mathrm{km\,s^{-1}}$ this event was clearly associated
with a mass ejection, while this question remains unsettled for DENIS~1048-39.

\subsection{Summary}

In conclusion, we find at least two active regions on DENIS~1048-39 contributing to the bulk
of the Balmer line flux. Mass motions directed towards the observer are found for the
emission lines of the Balmer series as well as for the Na D lines. In the last spectrum of our
observations the
onset of a flare seems to take place, since substantial brightening and blueshifts 
can be seen in the lines.
In addition in the $\mathrm{H_{\alpha}}$ and the $\mathrm{H_{\beta}}$ line there is a broad
emission feature on the blue side of the line. This can be interpreted as a rising  cloud.

Since DENIS~1048-39 seems to be located directly at the hydrogen burning limit this flare
gives evidence that such events may be more ubiquitous than previously assumed. It is consistent 
with X-ray detections of brown dwarfs \citep{Mokler} and the X-ray flare event found on the similar late-type star LHS~2065 \citep{LHS-2065}.



\bibliographystyle{aa}
\bibliography{papers}

\end{document}